\documentclass[AMA,STIX1COL, doublespace]{WileyNJD-v2}

\usepackage{savesym}
\usepackage{moreverb}
\savesymbol{bigtimes}

\usepackage{mathabx}
\restoresymbol{TXF}{bigtimes}
\usepackage{longtable}

\newcommand\BibTeX{{\rmfamily B\kern-.05em \textsc{i\kern-.025em b}\kern-.08em
T\kern-.1667em\lower.7ex\hbox{E}\kern-.125emX}}

\articletype{Article Type}%

\received{<day> <Month>, <year>}
\revised{<day> <Month>, <year>}
\accepted{<day> <Month>, <year>}

\begin{document}
\title{Efficient Study Design with Multiple Measurement Instruments}

\author[1]{Michal Bitan*}

\author[1]{Malka Gorfine}

\author[2]{Laura Rosen}

\author[1]{David M. Steinberg}

\authormark{AUTHOR ONE \textsc{et al}}

\address[1]{\orgdiv{Department of Statistics and Operations Research}, \orgname{Tel Aviv University}, \orgaddress{\state{Ramat Aviv}, \country{Israel}}}

\address[2]{\orgdiv{Department of Health Promotion, School of Public Health, Sackler Faculty of Medicine}, \orgname{Tel Aviv University}, \orgaddress{\state{Ramat Aviv}, \country{Israel}}}

\corres{*Michal Bitan, Department of Statistics and Operation Research, Tel Aviv University. \email{bennoac@post.tau.ac.il}}

\vspace{2cm}
\abstract[Abstract]{Outcomes from studies assessing exposure often use multiple measurements. In previous work, using a model first proposed by Buonoccorsi (1991), we showed that combining direct (e.g. biomarkers) and indirect (e.g. self-report) measurements provides a more accurate picture of true exposure than estimates obtained when using a single type of measurement. In this article, we propose a valuable tool for efficient design of studies that include both direct and indirect measurements of a relevant outcome. Based on data from a pilot or preliminary study, the tool, which is available
online as a shiny app \citep{shinyR}, can be used to compute: (1) the sample size required for a statistical power analysis, while optimizing the percent of participants who should provide direct measures of exposure (biomarkers) in addition to the indirect (self-report) measures provided by all participants; (2) the ideal number of replicates; and (3) the allocation of resources to intervention and control arms. In addition we show how to examine the sensitivity of results to underlying assumptions. We illustrate our analysis using studies of tobacco smoke exposure and nutrition. In these examples, a near-optimal allocation of the resources can be found even if the assumptions are not precise.}

\maketitle

\section{Introduction}
\label{sec1}
Scientists in various fields, including tobacco smoke exposure, nutrition, and environmental health  have struggled to accurately assess human exposure to various substances. In our previous research, we demonstrated that combining direct and indirect measures provides a more accurate picture of true exposure than does use of a single direct measure.  In this work we show how to exploit the earlier results to derive efficient study designs. With a given budget, our approach finds designs that obtain the highest power possible, and given a requirement on power, it finds the study design with the smallest possible budget.

Currently, multiple types of measurements  are often used in exposure studies. For example, both a biomarker (as a direct measure) and a self-report (as an indirect measure) may be included. Our research was motivated by a problem of study design for assessing tobacco-smoke exposure (TSE). TSE studies use biomarkers such as urinary cotinine, serum cotinine and hair cotinine, and also questionnaires on the frequency and intensity of exposure (Hovell et al \citep{hovell2009counseling}, Wilson et al. \citep{wilson2011randomized}, Kalkbrenner et al.\citep{kalkbrenner2010determinants}). Similar  measures are also used in nutritional research and clinical medicine. The TONE \citep{appel1995trial} study measured sodium intake, which was assessed both by urinary sodium and by analyzing answers to a 24-hour food recall questionnaire. A study \citealp{meirovitz2006grading} to reduce adverse effects (xerostomia) of radiation therapy in patients suffering from mumps and throat cancer included both salivary gland flow measurements (direct) and patient reports of symptoms. Similar problems arise in  other contexts. For example, in engineering, imaging methods (indirect) may be combined with occasional destructive measurements (direct) to assess internal properties such as crack length.\citealp{achenbach2000quantitative}

Designing such studies confronts us with a number of challenges. These include setting sample sizes, deciding what fraction of the participants should have a direct measurement, whether replicates are desirable and, when the study compares groups, what should be the allocation of resources to each group.  We answer those questions here in the context of a measurement error model that was introduced by Buonaccorsi \cite{buonaccorsi1991measurement}; however he did not address study design questions. Davidov and Haitovsky \cite{dabidov2000optimal}, working from a different model, considered a special case of the design problem that we address.  We discuss later how their results relate to ours.
We derive an explicit solution when all participants have direct measures and, for other cases, provide numerical solutions with a software tool that we developed and which is available online as a shiny app  \citep{shinyR}.  The tool can be used to find an optimal design for a known budget and to find the minimum budget design for a given constraint on power or on a confidence interval.  We present extensive results on the problem parameters that affect the choice of design and illustrate how to check the sensitivity of the resulting design to uncertainty in the parameters.

The article is organized as follows. In Section~\ref{sec:Methods}, we present the measurement error model for a single group study, the maximum likelihood estimator (MLE) for the mean, the standard error (SE) of the MLE, and a cost model for the study. Section~\ref{sec:studyDesign} presents the study-design calculations, and demonstrates the effect of the model and cost parameters on the design and on the SE. Section~\ref{sec:Extend}  extends the ideas to a two-group comparative study. Section~\ref{sec:settingsamplesize} covers power and sample size problems.  In Section \ref{sec:CaseStudy} three studies are used to illustrate the utility of our tool. We conclude with a short discussion in Section~\ref{sec:discussion}.

\section{The Model}
\label{sec:Methods}
We present here the measurement error model of Buonoccorsi \cite{buonaccorsi1991measurement} for a single-group study; the extension to two groups is straightforward.  Then we present the MLE for estimating the mean and its SE. For further details see \citep{talitman2020estimating}.  Finally we present a cost model for study design.

\subsection{The Measurement Error Model}\label{sec:ErrorModel}
Consider a set of $N$ independent individuals. Let $T_{j}$ denote the true level of participant $j$, $j=1,\ldots,N$. A simple model for $T_{j}$ is of the form
\begin{equation*} \label{Tij}
T_{j}=\mu+\epsilon_{j} \, ,
\end{equation*}
where $\mu$ is the population mean, $\epsilon_{j} \sim N(0,\sigma^2_{\epsilon})$, and $\sigma^2_{\epsilon}$ is the population variance. The main goal of the study is to estimate $\mu$.

We cannot observe $T_{j}$; however, we can observe various surrogates of it.  For example, TSE can be assessed by a biomarker (e.g. urinary cotinine) or by administering a questionnaire. We consider the former to be a direct, and unbiased, measurement of the true value $T_j$, whereas the latter is indirect and biased. We allow for the possibility of repeat direct measurements. We denote the direct measures by $M_{jk}$, $j=1,\ldots,n$, $k=1,\dots,K$, and model them by
\begin{equation*} \label{Mijk}
M_{jk}=T_{j}+\delta_{jk} \, ,
\end{equation*}
where $\delta_{jk}$ are independent and normally distributed with mean zero and a constant variance $\sigma^{2}_{\delta}$. Since direct measures could be expensive, we assume that these data may be available only for a sub-sample, so that $n \leq N$.  We also assume that all individuals in the sub-sample have the same number, $K$, of replicate direct measurements. Let the mean and variance of the replicates for the $j$th participant be denoted by $\widebar {M}_{j.}=\sum_{k=1}^{K}M_{jk}/K$ and $\sigma^{2}_{\widebar{M}_{j.}} = Var(\widebar{M}_{j.})=\sigma^{2}_{\epsilon}+\sigma^{2}_{\delta}/K$.
The setting with unequal replicates is described in detail in \citep{talitman2020estimating}.
We assume that the indirect measurement, which we denote by $Q_j$, is related to the true level via a regression equation and a classical measurement error model,
\begin{eqnarray*} \label{Qj}
Q_{j}=\alpha_{0}+\alpha_{1}T_{j}+\phi_{j} =\alpha_{0}+\alpha_{1}\mu+\alpha_{1}\epsilon_{j}+\phi_{j}
\end{eqnarray*}
where $\phi_{j} \sim N(0,\sigma^{2}_{\phi})$, and $\phi_{j}$ are independent of all other error terms.  We denote the population mean and variance of the indirect measurement by $\nu=E\left(Q_{j}\right)=\alpha_{0}+\alpha_{1}\mu$ and $\sigma^{2}_{Q}=Var(Q_{j})=\alpha^{2}_{1}\sigma^{2}_{\epsilon}+\sigma^{2}_{\phi}$, respectively.  For further details about the model see \citep{talitman2020estimating}.

We assume that all participants provide indirect measurements, which is standard practice in exposure studies with more than one measurement instrument.  However, direct measurement may be limited to a subset of the participants, who constitute a calibration sub-study.  For example, this might be desirable if direct measurement is much more expensive than self-reports.  It is sometimes useful to include replicates of indirect measures.  For example, 24-hour food recall questionnaires are often completed over several days, so that both weekday and weekend eating patterns are covered.  A questionnaire to assess TSE might include separate items like the number of cigarettes smoked near the participant in, and outside, the home; or by different individuals. When repeated indirect measures are observed, we assume that they will be combined into a single summary. These assumptions are natural in epidemiological exposure assessment, but they might not be appropriate for other metrology contexts.  We comment on these issues in the discussion.

\subsection{The MLE of $\mu$ and its Variance}
Talitman et al. \citep{talitman2020estimating} derived the MLEs for all the parameters in the model of section \ref{sec:ErrorModel}. We remark in passing that they also showed that the solution can easily be extended, via the EM algorithm, to obtain the MLE if the replication numbers $K_{j}$ are not equal; this will not be common as a design choice but may happen in practice due to missing data. They also obtained a closed-form expression for the variance of the estimated mean level $\mu$.

The MLE is described in  Appendix~\ref{sec:Appendix} . The important result for the design problem we study here is the variance formula,
\begin{equation*}
{\rm Var}\left(\widehat{\mu}\right)=\frac{\sigma^2_{\epsilon}}{Nn\left(n-3\right)}\left[\left(Nn-2N-n \right)\left(1+\frac{r_{\delta}}{K}\right)-\frac{\left(N-n\right)\left(n-2\right)}{1+r_{\phi}}\right] \, ,
\end{equation*}
where $r_{\phi}=\frac{\sigma^2_{\phi}}{\alpha^2_1 \sigma^2_{\epsilon}}$  and $r_{\delta}=\frac{\sigma^2_{\delta}}{\sigma^2_{\epsilon}}$.  The ratios compare the error variance in the indirect measure, after standardizing for the magnitude of the regression slope, and the variance of the direct measurement, to the variance of the population. For design purposes, prior values will be needed for the ratios, which could be based on a pilot study or data from related studies.

\subsection{The Cost Model}
We make the following budgetary assumptions:\\
A1. The study design is limited by an overall funding level $C$.\\
A2. The cost per individual of recruiting, maintenance, assessment and indirect measurement is $C_Q$.\\
A3. The cost for each direct measurement is $C_B$.\\
With $N$ participants and $K$ direct measurements on each of $n$ of them, the total cost  is $nKC_B+NC_Q $, which must be at most $C$. We develop our results using two ratios of the costs: $r_{CB}=\frac{C_B}{C_Q}$, which compares the cost of each direct measurement to that of recruitment, including an indirect measurement and $r_C=\frac{C}{C_Q}$, the cost of an indirect measurement relative to the total budget. Note that the latter ratio is the sample size if the study uses only indirect measurements. The corresponding constraint on the design parameters is $nKr_{CB} + N \leq r_C$.

\section{Study design} \label{sec:studyDesign}

We first consider the design problem for a study with a single group.  In the next section the results are extended to studies with two groups.

\subsection{The Design Problem}
Our goal is to find $N, n$ and $K$ that minimize ${\rm Var}(\widehat{\mu})$ given the budget constraint and the additional constraint that $n \leq N$.  Such designs will minimize the width of a confidence interval for $\mu$ and will maximize the power, $\pi$, for testing $H_0: \mu = \mu_0$.  If the true mean is $\mu$ and power is computed for a level $\alpha$ test, then
\begin{equation}\label{eq:powerOneGroup}
\pi \approx P\left[Z>\frac{z_{1-\frac{\alpha}{2}}-\left|\mu - \mu_0 \right|}{\sqrt{{\rm Var}\left(\widehat{\mu}\right)}}\right],
\end{equation}
where $Z$ is standard normally distributed. The approximation is due to ignoring the probability of rejecting the null hypothesis from a test statistic that gives a significant result ``in the wrong direction''. For purposes of sample size determination, researchers are interested in means and sample sizes for which that probability is negligible.

The design goal is clearly to minimize the standard error of the estimator in the denominator of Eq.~(\ref{eq:powerOneGroup}).
We first rewrite ${\rm Var}(\widehat{\mu})$ in terms of the cost ratios. Assuming that the budget constraint is exactly satisfied,
\begin{equation}\label{eq:varMu3}
{\rm Var}\left(\widehat{\mu}\right)=\frac{\sigma^2_{\epsilon}}{\left(n-3\right)}\left[\left(\frac{n-2}{n}-\frac{1}{r_C-nKr_{CB}} \right)\left(1+\frac{r_{\delta}}{K}\right)-\frac{n-2}{1+r_{\phi}}\left(\frac{1}{n}-\frac{1}{r_C-nKr_{CB}}\right)\right] .
\end{equation}
Using Eq.~(\ref{eq:varMu3}) and the estimates of the two ratios, $r_{\phi}$  and $r_{\delta}$, from a pilot study, we seek $N$, $n$ and $K$ to minimize ${\rm Var}\left(\widehat{\mu}\right)$.  In the following results, it is assumed that $\sigma^2_{\epsilon}=1$.

The optimization problem for $N,n$ and $K$  cannot, in general, be solved analytically. Therefore most of our results rely on numerical optimization using the genoud \citep{rgenoud} function in the R statistical environment, which is used in our shiny app \citep{shinyR}. Genoud is a function that uses evolutionary search algorithms to solve difficult optimization problems. Of particular relevance for our problem is that genoud can optimize integer parameters.

\subsection{What Parameters Reduce the Variance?}

\begin{figure}[p]
\begin{center}
\includegraphics[scale=0.75]{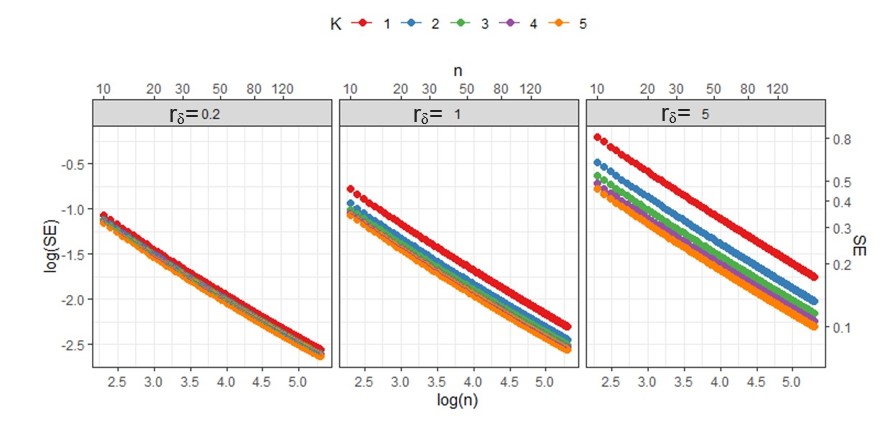}
\caption{SE as a function of the sample size, the number of repeated direct measurements and $r_{\delta}$ (0.2, 1 or 5).}
\label{Fig:SEnk}
\end{center}
\end{figure}

Before presenting results on efficient design, we first give some understanding of
the behaviour of the SE as a function of $n$ and $K$ without the influence of the costs. We set $N=200$ and varied $K$ to be $1, \dots, 5$ and $n=10,\ldots,200$. We varied the ratios $r_{\phi}$ and $r_{\delta}$ to be $0.2, 1$ or $5$.  We found that $r_\phi$ has very little effect on the SE.  Consequently, the results, which are shown in Figure~\ref{Fig:SEnk}, are for the case $r_\phi=1$. The figure plots the SE against $n$ on a log scale and shows a very nearly linear relationship for all the variance ratios we considered. It can be seen from the figure that as $r_{\delta}$ decreases, the SE decreases.  When $r_\delta$ is small, $K$ hardly affects the SE,  but as $r_\delta$ increases, $K$ has a larger effect.  If $n=N$, then the SE decreases exactly at rate $1/\sqrt {n}$.  However, in the general case, examination of the results in Figure \ref{Fig:SEnk} shows that the SE decreases at a slightly slower rate than $1/\sqrt {n}$. When $n < N$ the benefit from participants with indirect measures only leads to a rate that is slower than $1/\sqrt {n}$ because
there is a lot of information not just from the direct measures.

\subsection{How Do the Parameters Affect the Optimal Design?}
\subsubsection{Optimal choice of K} \label{sec:optimalK}

When is it desirable to replicate the direct measurements?  We begin with the special case where other considerations dictate that direct measurements will be obtained from all participants, i.e. that $n=N$. In that case, the primary design decision is to fix $K$, the number of replicate direct measurements.  The variance function of $\mu$ is
\begin{equation*}\label{eq:varMuNn}
{\rm Var}\left(\widehat{\mu}\right)=\frac{\sigma^2_{\epsilon}}{N}\left(1+\frac{r_{\delta}}{K}\right)=\frac{\sigma^2_{\epsilon}}{C}\left(\frac{C_Qr_\delta}{K}+KC_B+C_Q+C_Br_\delta\right).
\end{equation*}
Comparing the variances when $K=k-1$ and $K=k$, we find that $K=k$ is the better choice when
\begin{equation*} \label{eq:condK}
k(k-1)< \frac{r_\delta}{r_{CB}} \, .
\end{equation*}
Hence, the optimal choice of $K$ will be the largest value for which the above inequality holds.
The solution when $n=N$ provides useful intuition.  It is beneficial to take more direct measurements when (1) the direct measurements are noisier, and thus less informative, and (2) the direct measurements are cheap relative to the cost of recruitment and indirect measurement.

We assess the general case numerically.  Empirical evidence indicates that, as in the $n=N$ case, larger values of $r_\delta$ result in larger values for the optimal choice of $K$.  It is useful to focus on the question:  how large must $r_\delta$ be to require more replicates, comparing $K=2$ to $K=1$ and $K=3$ to $K=2$.  These comparisons were carried out for a variety of settings of the cost parameters; we set $C=2,000,000$ and varied $r_{CB}$  to be $0.05, 0.1, 0.2, 0.25, 0.4, 0.5, 1, 2, 2.5, 4, 5, 10$ or $20$.  Because we found that $r_{\phi}$ has a small impact on the SE , we set $r_{\phi}=1$.

\begin{figure}[p]
\begin{center}
\includegraphics[scale=0.75]{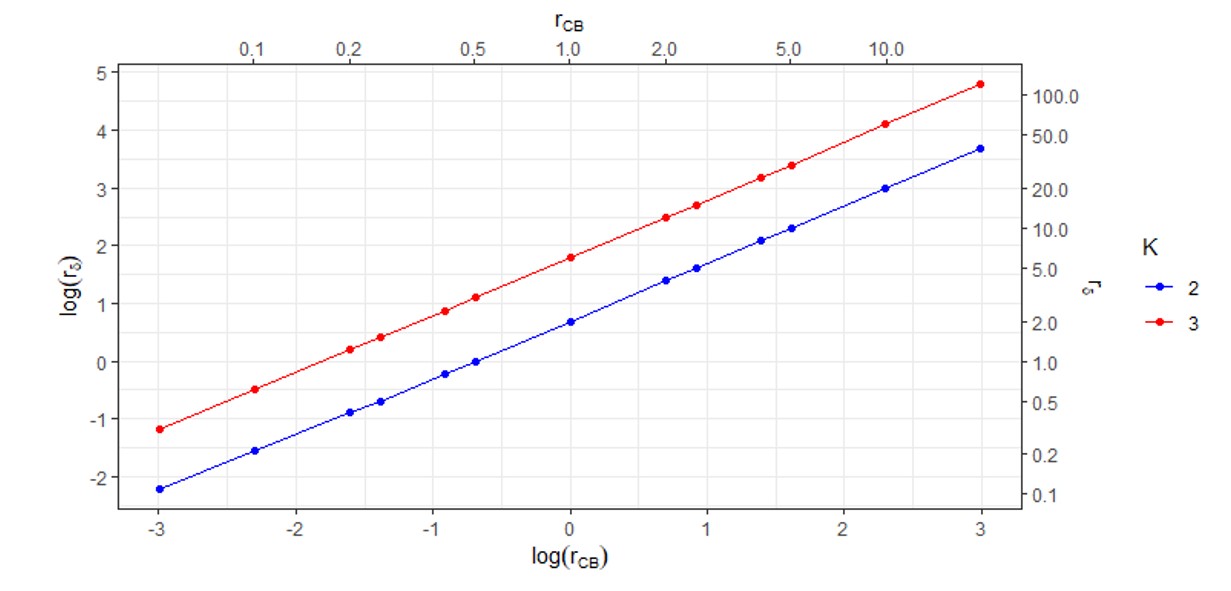}
\caption{Optimal $K$ according to $r_\delta$ and $r_{CB}$. Below the blue line $K=1$ is better, above the red line $K \geq 3$ is better and between the two lines $K=2$ is better.}
\label{Fig:whenKbetter}
\end{center}
\end{figure}

The results are shown in Figure \ref{Fig:whenKbetter}, which is a log-log plot of the threshold values of $r_{\delta}$ as a function of $r_{CB} = C_B/C_Q$.  The thresholds are almost perfectly linear and the results are remarkably consistent across the range of $C_B$ and $C_Q$ values that were examined. The fitted regression lines $\log (r_\delta)=a+b\log (r_{CB})$ for the two comparisons are as follows:  the border between $K=1$ and $K=2$ is $\frac{r_\delta}{r_{CB}}=2.02 $ and $K=3$ versus $K=2$ is $\frac{r_\delta}{r_{CB}}=6.01$. These boundaries correspond almost exactly to the conditions for preferring $K=2$ to $K=1$, or $K=3$ to $K=2$, in the case that $n=N$.

The empirical results show that, over a wide range of realistic settings, the theoretical results on the optimal choice of $K$ (for $K = 1,2,3)$ in the special case $n=N$ continue to be excellent guidelines. We conjecture that the theoretical result in the case that $n=N$ will continue to be approximately valid for comparing larger values of $K$.

\subsubsection{How does $\frac{n}{N}$ relate to the parameters?}
\begin{figure}
\begin{center}
\includegraphics[scale=0.75]{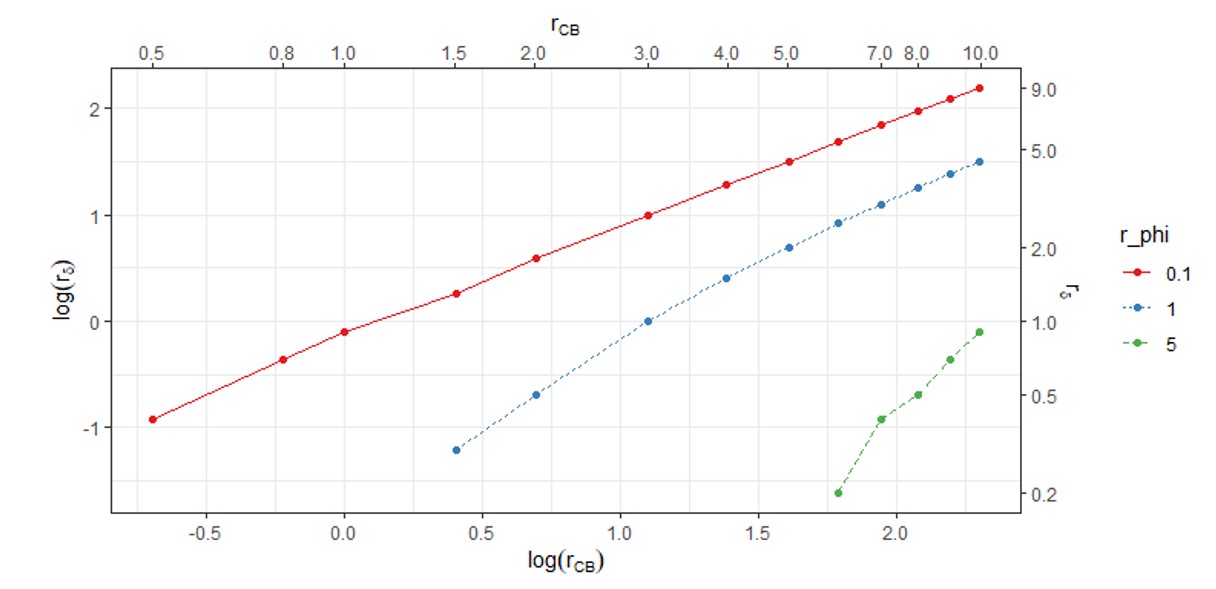}
\caption{The optimal sampling fraction $\frac{n}{N}=1$ above the lines relating $r_\delta$ to $r_{CB}$.  Each line corresponds to a different value of $r_\phi$. For all settings shown, the optimum $K$ is $1$.}
\label{Fig:whenn_N_1_loglog}
\end{center}
\end{figure}

Here we demonstrate how the optimal fraction of participants with direct measurements, $\frac{n}{N}$, is affected by the other parameters. We begin by investigating when it is desirable to include direct measurements for all the participants. Across a wide range of values for the other parameters, the optimal fraction is 1 when $r_\delta$ and $r_{CB}$ exceed certain thresholds. Numerically, we determined the minimal value of $r_\delta$ or the maximum value of $r_{CB}$ that gives $\frac{n}{N}=1$ when all other parameters are fixed. Figure \ref{Fig:whenn_N_1_loglog} shows regions in the $r_\delta$ by $r_{CB}$ plane for which the optimal sampling fraction equals $1$, for several values of $r_{\phi}$. In all the cases shown in Figure \ref{Fig:whenn_N_1_loglog}, the optimal $K$ was $1$. The constraints sometimes lead to solutions for which $n < N$ only because the overall budget is sufficient for additional recruitment but not for additional direct measurement. The reduction in variance from these final participants was negligible. We record the optimal fraction in these cases to be 1.

\begin{figure}
\begin{center}
\includegraphics[scale=0.75]{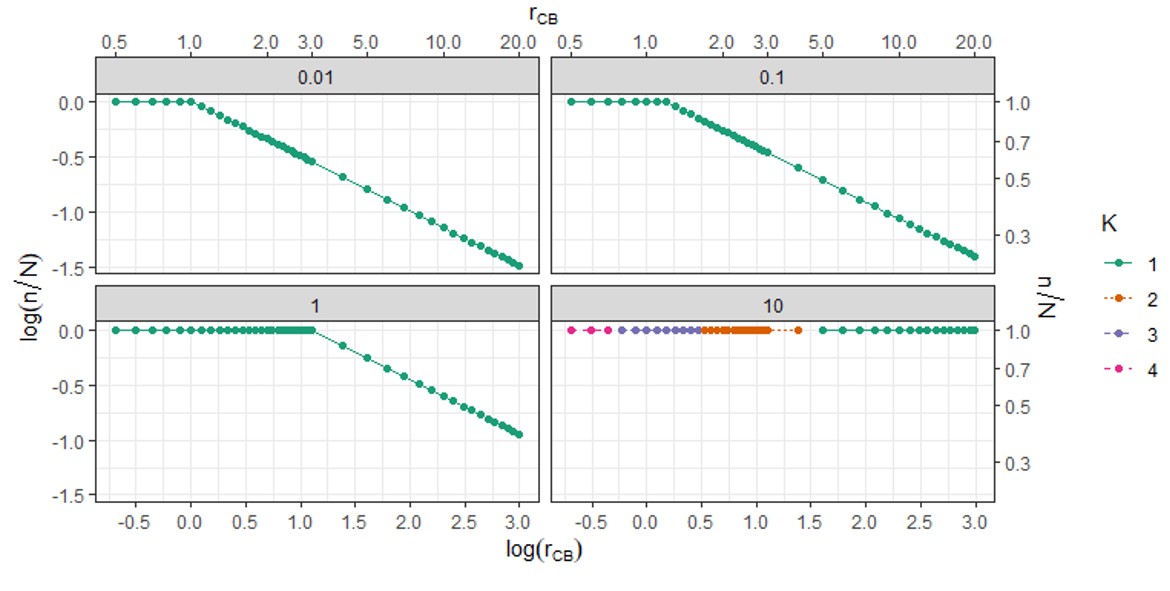}
\caption{The optimal $\frac{n}{N}$ and $K$ (indicated by color) as a function of $r_\delta$ (0.01, 0.1, 1 or 10) and $r_{CB}$ when $r_\phi=1$. }
\label{Fig:n_NOptimK}
\end{center}
\end{figure}

\begin{figure}
\begin{center}
\includegraphics[scale=0.75]{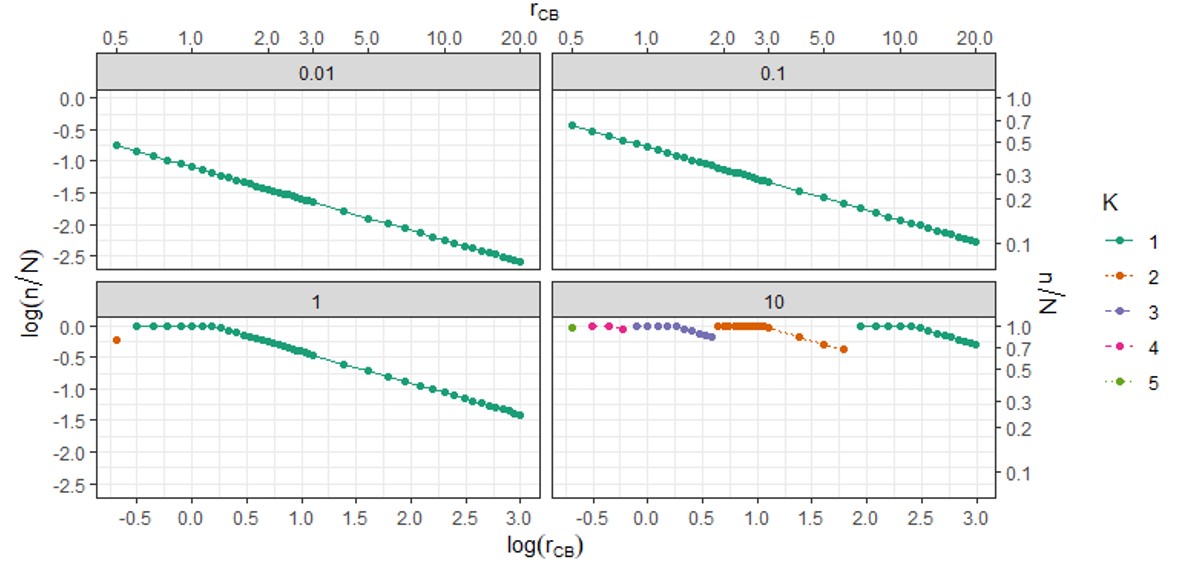}
\caption{The optimal $\frac{n}{N}$ and $K$ (indicated by color) as a function of $r_\delta$ (0.01, 0.1, 1 or 10) and $r_{CB}$ when $r_\phi=0.1$. }
\label{Fig:n_NOptimK0.1}
\end{center}
\end{figure}

In general, we see from Figure \ref{Fig:whenn_N_1_loglog}, as expected, that $\frac{n}{N}$ is an increasing function of $r_\delta$ and $r_\phi$ and a decreasing function of $r_{CB}$.
To understand how the ratio $\frac{n}{N}$ is related to the other parameters, we examined the following combinations: $r_\delta=0.01, 0.1, 1, 10, r_\phi=0.1, 1, r_{CB}=0.5, 0.6,0.7,\ldots,3,4,5,\ldots,20$ and $C_Q=1$.
The results are shown in Figures \ref{Fig:n_NOptimK} and \ref{Fig:n_NOptimK0.1}.
When the measurement error variance ratios $r_{\delta}$ and $r_{\phi}$ are small and the cost ratio $r_{CB}$ is high, the optimal $n/N$ can be small; such cases were shown in Davidov and Haitovsky \citep{dabidov2000optimal}.  However, the sampling fraction is not strictly monotone in either $r_{CB}$ or $r_\delta$.  When certain thresholds are crossed, the optimal $K$ changes and the optimal fraction can change direction at those points.  The non-monotone relationship can be seen in the lower right panel of Figure \ref{Fig:n_NOptimK0.1}.  As $r_{CB}$ increases, the optimal value of $K$ decreases.  At each threshold value where this occurs, the decrease in $K$ is accompanied by an increase in $n/N$.

\section{Extension to a two-group study}\label{sec:Extend}
It is straightforward to extend the above results, on inference for a single group, to a comparative study in which individuals have been randomized to either a control group or to an intervention, with $N_{i}$ individuals in group $i$, $i=1,2$.
We again assume that the study obtains an indirect measurement of outcome for each participant and a direct measurement for a subset of $n_i$ individuals. The focal point of interest is the intervention effect $\mu_2-\mu_1$. The design goal is to select
$N_i, n_i$ and $K_i$, $i=1,2$, to maximize the power for a desired effect size $\mu_2-\mu_1$.  The power is given by
\begin{equation}\label{eq:power}
\pi \approx P\left[Z>\frac{z_{1-\frac{\alpha}{2}}-\left|\mu_2 - \mu_1 \right|}{\sqrt{{\rm Var}\left(\widehat{\mu}_{1}\right)+{\rm Var}\left(\widehat{\mu}_{2}\right)}}\right],
\end{equation}
with the approximation from ignoring the probability of rejection due to a strong result in the wrong direction.

The power is maximized by minimizing the SE in the denominator of Eq.~(\ref{eq:power}).  For any division of resources between the two groups, that entails minimizing the variance within each group, given their resources. Hence, the study-design problem can be  partitioned into two parts: determining the resource allocation between the groups and finding the optimal study design for each group, given the allocation.  The latter problem is the one-group optimization that was discussed in Section \ref{sec:studyDesign}.  The former involves just one parameter and can be solved numerically

If both groups have the same population variance, that parameter does not affect the design decisions. If the variances differ, they affect only the allocation of resources between the two groups.
That allocation is also affected by the two ratios, $r_\delta$ and $r_\phi$, for each group.

\begin{figure}
\begin{center}
\includegraphics[scale=0.75]{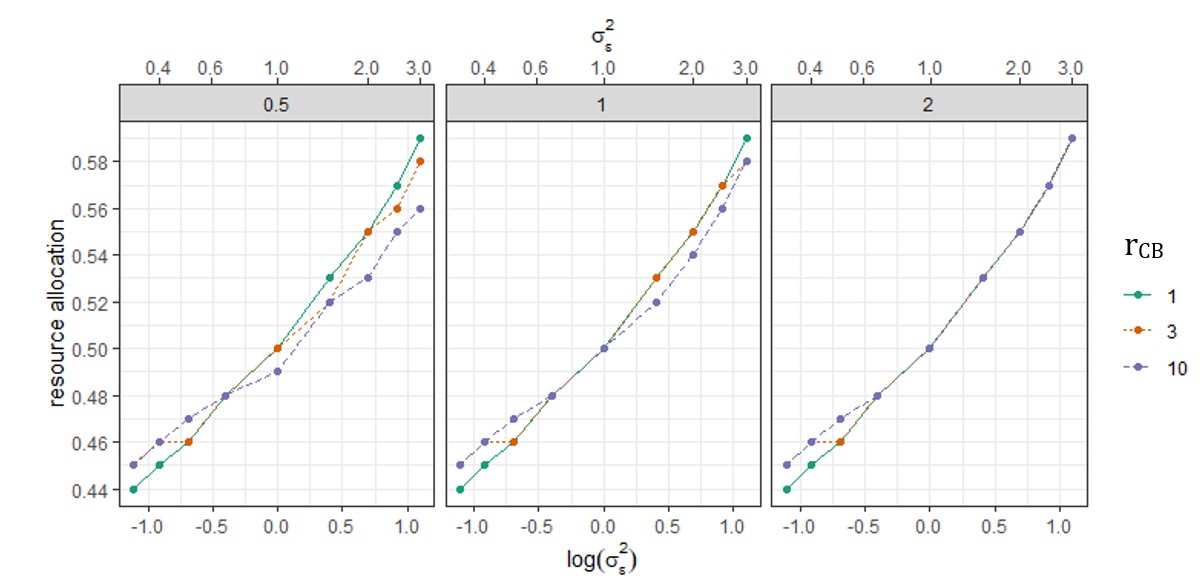}
\caption{Resource allocation for group 1 as a function of $r_{CB}, \sigma^2_\epsilon$ and $r_\phi$ (0.5,1,2) for that group.}
\label{Fig:allocation_log}
\end{center}
\end{figure}

As before, numerical optimization provides insight into the choice of design.  We start by studying how the resource allocation is affected by $r_{\phi}$ and $\sigma^2_{\epsilon}$, assuming that $\sigma^2_\delta$ is equal in the two groups.  This will be reasonable for many applications, where $\sigma^2_\delta$ is largely determined by the precision of the measurement process, with the same process applied to both study groups.  To compare the allocations, we set the parameters $r_{\phi}$ and $\sigma^2_{\epsilon}$ to be $1$ in group $2$, changing the parameters in group $1$ as follows:
$r_\phi=0.5, 1, 2$ and $\sigma^2_\epsilon=0.33, 0.4, 0.5, 0.667, 1, 1.5, 2, 2.5, 3$. The fraction of resources devoted to group $1$ is shown in Figure \ref{Fig:allocation_log}. The allocation is largely a function of $\sigma^2_\epsilon$, with more resources allocated to the group with higher variance. The ratios $r_{CB}$ and $r_\phi$ have almost no effect on the allocation.

\section{Setting sample size}\label{sec:settingsamplesize}

The results of the previous sections can be used to guide choice of sample size.  There are two well-known approaches:  (\textit{i}) to meet demands on power for testing a null hypothesis and (\textit{ii}) to achieve a confidence interval of a desired width.  Both approaches lead to conditions on the standard error of the effect estimator.  We provide details for the case of hypothesis testing in a two-group study with a two-sided $H_0$.  The same ideas can easily be applied to one-sample problems, one-sided hypotheses, and to the confidence interval approach.  The sample size for testing $H_0$ requires a pre-specified significance level $\alpha$, power $\pi$ and expected effect size of $\mu_2-\mu_1$.  The target standard error of the effect estimator must then satisfy the inequality
\begin{equation}\label{eq:se-target}
SE_{target} \leq \frac{\left|\mu_2-\mu_1\right|}{Z_{1-\alpha/2}+Z_{\pi}}.
\end{equation}

In standard problems, enforcing equality in \ref{eq:se-target} leads directly to a formula for the sample size.  In our calibration sub-study setting, the problem is to minimize the budget $C$.  Denote by $SE(C)$ the standard error achieved for a budget $C$ by the study design with optimal choices of $N$, $n$ and $K$.  Trivially, $SE(C)$ is a decreasing function of $C$, but the relationship is too complex, in general, to provide a simple inversion formula to compute the value of $C$ that gives $SE_{target}$.
A simple iterative scheme can be used to solve for $C$.

The scheme begins by assuming that $n_i = N_i$ for both groups and that half the resources are allocated to each of the groups.  These are the assumptions made in section \ref{sec:optimalK}. We further assume that the guidelines in that section give the optimal choices of $K_i$.  Applying equation \ref{eq:varMuNn} for both groups now gives an equation in which the standard error is proportional to $C^{-0.5}$.  Equating this expression to $SE_{target}$ gives the initial solution
\begin{equation}
C_0=\frac{2\sigma_\epsilon^2}{SE^2}\left \{ \left(\frac{C_Qr_{\delta_1}}{K_1}+K_1C_B+C_Q+C_Br_{\delta_1} \right)^2+\left(\frac{C_Qr_{\delta_2}}{K_2}+K_2C_B+C_Q+C_Br_{\delta_2} \right)^2 \right \}
\end{equation}
Find the optimal design for $C_0$ and compute $SE(C_0)$.  If $SE(C_0) = SE_{target}$, the design and budget have been found.  However, the design used to compute $C_0$ may not be optimal, so that $SE(C_0) < SE_{target}$.  In that case, iteratively correct the budget by $C_{i+1} = (SE(C_i)/SE_{target})^2C_i$.  The formula is based on the assumption that, in the region of $C_i$, $SE(C)$ will still be proportional to $C^{-0.5}$, but perhaps with a different proportionality constant than the one assumed when computing $C_0$.
In a number of examples we have found that this method converges very rapidly, even with a small convergence tolerance of, say, $0.001\%$ difference from the target.

\section{Case study}
\label{sec:CaseStudy}
In this section we illustrate our ideas for study design using three published intervention trials. Two of them (Hovell et al.\citep{hovell2009counseling} and Wilson et al.\citep{wilson2011randomized}) examined programs to reduce children's exposure to tobacco smoke and encourage parental smoking cessation. Both studies used the biomarker urinary cotinine as a direct measure of exposure and parental self-reports as the indirect measure. The third example is the Trial of Nonpharmacologic Intervention in the Elderly (TONE), which assessed the effects of weight loss or reduction in sodium intake, or both, on blood pressure control in individuals who were taken off antihypertensive medication. Full details of the trial are given by Appel et al \cite{appel1995trial}. We focus here on participants' sodium levels, which were measured directly by urinary sodium and indirectly from 24-hour food recall questionnaires. All three studies included replicates of the direct measurements. The TONE study had $2$ replicates and Wilson's and Hovell's studies had $3$. Maximum likelihood estimates of the model parameters for these studies are shown in Table \ref{tbl:initParam}. They cover a wide range of settings in terms of the variance ratios.  Both smoke exposure studies used the same biomarker, yet found substantially different levels of ``within participant'' variation across replicates.  This suggests that careful assessment may be needed at the pre-study planning phase that we address here.

We express design comparisons in terms of ``efficiency,'' defined as the ratio of the smaller variance to the larger variance.  In general, this can be translated into cost differences: if design A is 50\% efficient compared to design B, the team would need to double the budget for design A to achieve the same variance as design B.

Our cost figures are based on input from researchers in the field of tobacco smoke exposure alongside published information. We assume that recruitment, including the self-report questionnaire, will cost between $\$50 $ and $\$150 $ per participant \cite{renwick2018cost}, that each urinary cotinine measurement will cost between $\$150 $ and $\$250 $ and each urinary sodium measurement will cost $\$250 $ \cite{ninci2004much}. We also assume that the measurement variance of the biomarker, $\sigma_{\delta}^2$, is identical for both groups.

The study-design question then has two components, first to determine the resource allocation between the two groups, and second to determine the optimal study design within each group. The results for each study are shown in Table \ref{tbl:designRes}.
We compared two total budgets, one close to what the researchers used in their studies ($ \$50,000$) and the other much higher ($ \$250,000$).
For Hovell and TONE, increasing the budget by a factor of $5$ simply increases all the sample sizes by a factor of $5$.  However, this need not always happen, as illustrated by Wilson's data, where the higher budget makes it advantageous to remove the replication in group $1$, so that the sample size there increases by a factor of about $8$. If we force $K$ to be the same in each group, e.g. to $K=1$, the design is only slightly worse, with an efficiency of $98.6\%$; so one could easily justify either choice of $K$.
In all the cases, more resources are given to the group with the higher population SD. This matches what was seen in Figure \ref{Fig:allocation_log}.

\begin{figure}
\begin{center}
\includegraphics[scale=0.75]{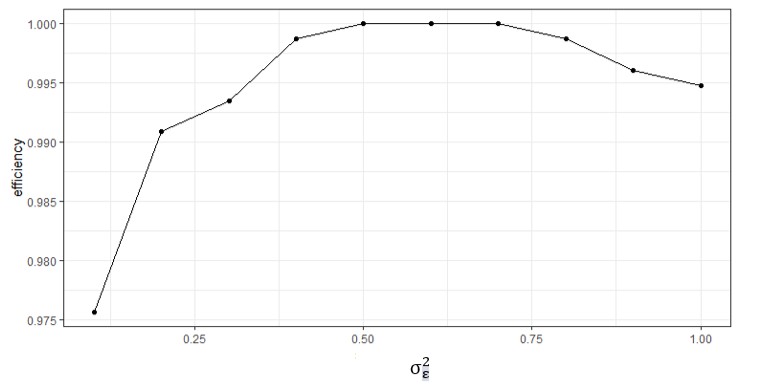}
\caption{Design efficiency as a function of $\sigma_\epsilon$ when $r_{CB}=2$ using parameter estimates from Hovell's study.}
\label{Fig:hovellEfficiencysig}
\end{center}
\end{figure}

\begin{figure}
\begin{center}
\includegraphics[scale=0.75]{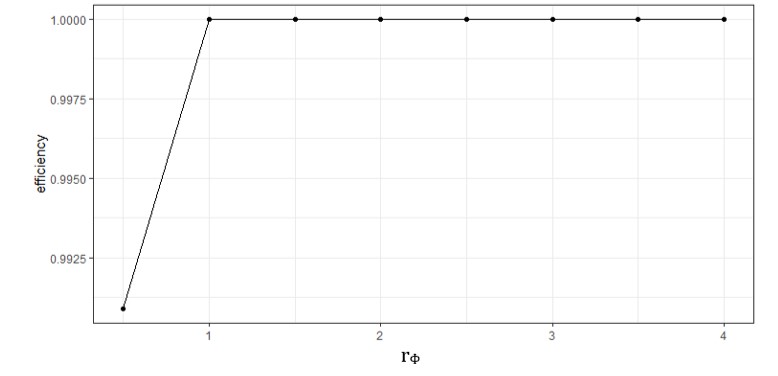}
\caption{Design efficiency as a function of $r_\phi$ when $r_{CB}=2$ using parameter estimates from Hovell's study.}
\label{Fig:hovellEfficiency_rphi}
\end{center}
\end{figure}

Researchers will often be unsure of the correct values of the input parameters, so it is informative to assess the sensitivity of the study design to inaccurate values at the planning stage.
We assume accurate cost assessment and consider how poor prior guesses of $\sigma^2_\epsilon$ and $r_\phi$ affect design efficiency.  We assume that these parameters are assessed correctly in group 2, but not in group 1, varying the values adopted at the design phase about the true values by up to a factor of 2.  As above, we take as the true values the estimates in Table \ref{tbl:initParam}.
Similar results were found for all studies, which we illustrate using Hovell's study.  Figure \ref{Fig:hovellEfficiencysig} shows efficiency as a function of the planning value for $\sigma^2_\epsilon$ and Figure \ref{Fig:hovellEfficiency_rphi} as a function of $r_\phi$. Efficiency is defined as the ratio of the variance with the true parameters to the variance with the assumed parameters. In all the settings examined, the efficiency remains above 0.975, suggesting that precise knowledge of the parameters is not crucial with respect to finding a nearly-optimal allocation of the resources.  In all three studies we found, as seen in Figure \ref{Fig:hovellEfficiencysig}, that under-assessment of $\sigma^2_\epsilon$ leads to a somewhat greater loss of efficiency than does over-assessment.
Inaccurate assessment of $r_\phi$ had no effect at all on efficiency in the Wilson and TONE studies; in Hovell's study, there was a very small decrease in efficiency only when the planning value of $r_\phi$ was lower than the true value.

\begin{figure}
\begin{center}
\includegraphics[scale=0.75]{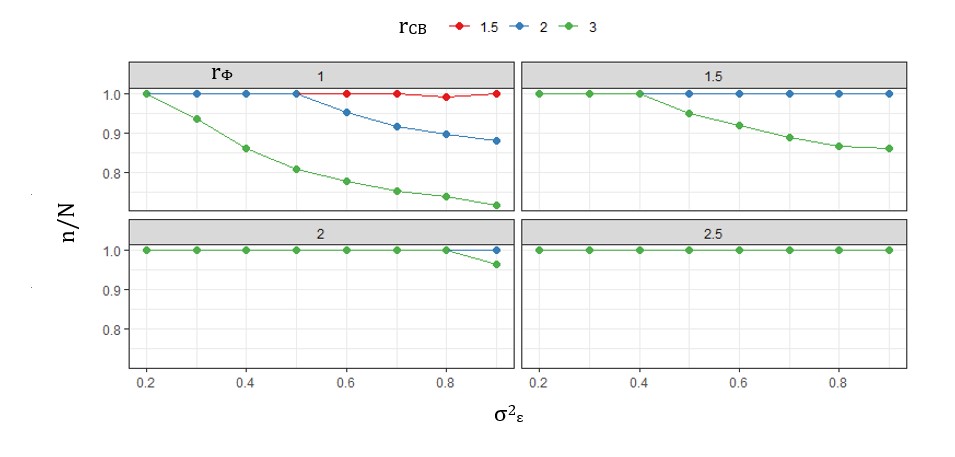}
\caption{The optimal sampling fraction $\frac{n}{N}$ as a function of the ratio of the budget $r_{CB}$, $r_\phi$ and $r_\delta$ using parameter estimates from Hovell's study.}
\label{Fig:Hovell1}
\end{center}
\end{figure}

In both sensitivity tests following the Wilson and TONE data, the ratio $\frac{n}{N}$ was close to one for all settings examined. For Hovell's data, a planning value of $r_\phi > 2$ leads to designs in which the researcher takes direct measures from all participants (see Figure \ref{Fig:Hovell1}). However, when the planning $r_\phi <= 1.5$, and $\sigma^2_\epsilon$ and $r_{CB}$ are sufficiently large, an increasing fraction of participants should have only indirect measures.

The sensitivity analysis for the Hovell and TONE studies consistently finds that the optimal value of $K$ is $1$, but for Wilson's data, the optimal $K$ ranged from $1$ to $4$. This difference is driven by $r_{\delta}$, which is much higher in Wilson's study than in Hovell's study or in the TONE data.

Table \ref{tbl:designRes2} illustrates how the design changes when modifying the required power from 80\% to 90\%, for $\mu_2-\mu_1=0.1$, and $\alpha=0.05$, using the parameters from Hovell's study. As the power increases, the $SE_{target}$ decreases and the budget must be increased.

\section{Discussion}\label{sec:discussion}

In studies that use both direct and indirect measurements, efficient design depends on several input parameters: $r_{CB}$, the ratio of the cost of a direct measurement to that of an indirect measurement; $r_{\delta}$, the ratio of the variance of the direct measurement to the variance of the population; and $r_{\phi}$, the ratio of the error variance in the indirect measure to the variance of the population, after standardizing for the magnitude of the regression slope that relates the indirect measurement to the true value.

The problem inputs $r_{\delta}$ and $r_{CB}$ are related to the relative benefit of taking replicate direct measurements: when $r_{\delta}/r_{CB}$ is small, the study should make only one direct measurement per participant, but as their ratio increases, it becomes beneficial to add replicates.  We found a simple condition for setting the optimal number of replicates $K$.

Optimal designs do not necessarily require direct measurements from all participants.  The ratios $r_{\delta}$ and $r_{CB}$ are again key inputs for guiding the fraction that gets direct measurements.  As with the number of replicates, the sampling fraction $n/N$ is increasing in $r_{\delta}$ and decreasing in $r_{CB}$.  However, there is also interplay with the optimal number of replicates:  these same trends affect the optimal number of replicates, and when that number increases, it breaks the monotonicity of the relationships to $r_{\delta}$ and $r_{CB}$. The variance ratio for the error in the indirect measurements, $r_{\phi}$, also affects the sampling fraction.  These results are consistent with, and considerably expand on, designs presented by Davidov and Haitovsky \cite{dabidov2000optimal}.  They did not consider the option of replicate direct measurements and focused on settings with quite high correlation between the direct and indirect measurements, much higher than what we have seen in exposure or nutritional studies.

With two groups, the allocation of resources between them depends primarily on their relative population variances, with more resources, and hence larger sample sizes, for the group that
has larger variance.

Using statistical criteria to set sample size is a standard concern in study design, and usually a requirement of funding bodies evaluating research proposals.  These criteria lead to constraints on the SE of effect estimates which, in simple problems, translate directly into bounds for sample size.  We show here that in research that exploits both direct and indirect measurements, sample size determination is more complex, involving both the relative precision and the relative costs of the two types of measurements.  Constraints on the SE translate into bounds on the overall budget.  The specific design parameters are then derived implicitly by finding the optimal design for the minimal budget.  We provide a framework and a software tool that allows research teams to solve these problems.

There are a number of interesting extensions.  Many exposure studies examine long-term effects of interventions by collecting longitudinal data on participants.  Primary outcomes are usually based on comparison of outcomes at baseline and at the study termination.  Our approach can be used to determine optimal policies for deciding which subjects, and at which time points, should be measured directly.  The study will often include intermediate time points, as well, and a natural concern is whether direct measurement is required at all time points or only at baseline and conclusion. As pointed out by the referees, another useful extension is to include subject-level covariates in our measurement models. We plan to address these problems in subsequent work.

\section{Supplementary Material}
The reader is referred to the on-line Supplementary Materials for technical appendices.

\section*{Acknowledgements}
We gratefully acknowledge the support from the Israel Science Foundation (ISF 1067/17), which supported M.B. and M.G. and the support from the Flight Attendant Medical Research Institute (FAMRI), which provided partial support to M.B. and L.R.

\bibliography{SSizerefs}

\begin{table}[p]
\caption{Parameter estimates for the Hovell, Wilson and TONE studies}
\centering
\begin{tabular}{|c|c|c|c|c|c|c|}
\toprule
 & \multicolumn{2}{|c|}{Hovell} & \multicolumn{2}{|c|}{Wilson} & \multicolumn{2}{|c|}{TONE}\\
\midrule
 & Estimate & SE & Estimate & SE & Estimate & SE \\
\hline
$\widehat{\alpha}_{01} $ & 1.630 & 0.481 & 2.109 & 0.342 & -0.158 & 0.751 \\
$\widehat{\alpha}_{11} $ & 0.840 & 0.168 & 0.126 & 0.146 & 0.898 & 0.162 \\
$\widehat{\sigma}^2_{\epsilon_{1}} $ & 0.551 & 0.113 & 0.778& 0.185& 0.113 & 0.018 \\
$\widehat{\sigma}^2_{\phi_{1}} $ & 0.692 & 0.215 & 0.846 &0.102 & 0.289 & 0.026 \\
$\widehat{\sigma}^2_{\delta} $ & 0.237 & 0.031 & 3.072 & 0.266 & 0.225 & 0.015 \\
$\widehat{\alpha}_{02} $ & 1.729 & 0.322 & 2.128 & 1.280 & 1.748 & 0.332 \\
$\widehat{\alpha}_{12} $ & 0.868 & 0.139 & 0.120 &0.483 & 0.442 & 0.080 \\
$\widehat{\sigma}^2_{\epsilon_{2}} $ & 0.705 & 0.122 & 0.486 & 0.203 &0.210 & 0.024 \\
$\widehat{\sigma}^2_{\phi_{2}} $ & 0.740 & 0.175 &0.685 & 0.109 & 0.284 & 0.019 \\
$\widehat{r}_{\phi_{1}} $ & 1.78 & & 64.48 & & 3.26 & \\
$\widehat{r}_{\delta_{1}} $ & 0.43 & & 3.95 & & 1.99  &\\
$\widehat{r}_{\phi_{2}} $ & 1.40 & & 96.37 & & 6.89  & \\
$\widehat{r}_{\delta_{2}} $ & 0.34 & & 6.32 & & 1.07 &\\
\hline
$N_1$ & \multicolumn{2}{|c|}{63} & \multicolumn{2}{|c|}{109} & \multicolumn{2}{|c|}{420}\\
$n_1$ & \multicolumn{2}{|c|}{63}& \multicolumn{2}{|c|}{109}& \multicolumn{2}{|c|}{420}\\
$K_1$ & \multicolumn{2}{|c|}{3}& \multicolumn{2}{|c|}{3}& \multicolumn{2}{|c|}{2}\\
$N_2$ & \multicolumn{2}{|c|}{68}& \multicolumn{2}{|c|}{123}& \multicolumn{2}{|c|}{430}\\
$n_2$ & \multicolumn{2}{|c|}{68}& \multicolumn{2}{|c|}{123}& \multicolumn{2}{|c|}{430}\\
$K_2$ & \multicolumn{2}{|c|}{3}& \multicolumn{2}{|c|}{3}& \multicolumn{2}{|c|}{2}\\
\bottomrule
\end{tabular} \label{tbl:initParam}
\end{table}

\begin{table}[h]
\caption{Study design results based on the Hovell, Wilson and TONE studies. Ratio is the fraction of resources allocated to group $1$. }
\centering
\begin{tabular}{cccccccc}
\toprule
& \multicolumn{3}{c} { $C=\$50,000$} & \multicolumn{3}{c} {$C=\$250,000$} \\
\midrule
 & Hovell & Wilson & TONE & Hovell & Wilson & TONE\\
 \hline
$n_1$ & 64 & 40 & 61 & 320 & 340 & 313 \\
$N_1$ & 64 & 40 & 62  & 320 & 340 & 314\\
$K_1 $ & 1 & 2 & 1 & 1 & 1 & 1 \\
$n_2 $ & 69 & 40 & 72 & 346 & 196 & 353 \\
$N_2 $ & 70 & 40 & 72 & 348 & 196 & 354\\
$K_2 $ & 1 & 2 & 1 & 1 & 2 & 1\\
Ratio & 0.48 & 0.50 & 0.46 & 0.48 & 0.51 & 0.47 \\
\bottomrule
\end{tabular} \label{tbl:designRes}
\end{table}

\begin{table}[h]
\caption{Design results based on Hovell's data as a function of $\pi$, $\mu_2-\mu_1=0.1$ and $\alpha=0.05$. \\Ratio is the fraction of resources allocated to group $1$.  \\ The tolerance = $0.00001$. The final budget is $C=C_1$. }
\centering
\begin{tabular}{ccc}
\toprule
&  $\pi=0.8$ & $\pi=0.9$ \\
\midrule
$C$ & 1,016,565 & 1,360,757 \\
$n_1$ & 1,301 & 1,741 \\
$N_1$ & 1,301 & 1,741 \\
$K_1 $ & 1 & 1 \\
$n_2 $ & 1,409 & 1,886 \\
$N_2 $ & 1,409 & 1,886 \\
$K_2 $ & 1 & 1 \\
Ratio & 0.48 & 0.48  \\
SE & 0.03569 & 0.03082 \\
$SE_{target}$ & 0.03569 & 0.03085\\
$SE(C_0)$ & 0.03566 & 0.03082 \\
$SE(C_1)$ & 0.03569 & 0.03085 \\
$C_0$ & 1,018,393 & 1,363,339\\
number of iterations & 1 & 1 \\
\bottomrule
\end{tabular} \label{tbl:designRes2}
\end{table}

\newpage

\section{Appendix}
\label{sec:Appendix}
We present here further details about the model and the MLE method for estimating the intervention effect.

In order to present the MLE, we need to repeat a few of the ideas in Talitman et al. \citep{talitman2020estimating}.  The model assumes a regression relationship of $Q_j$ to $T_j$, which immediately implies a regression of $Q_j$ on $\widebar{M}_{j.}$, with the same coefficients $\alpha_0$ and $\alpha_1$. In turn, Talitman et al. \citep{talitman2020estimating} showed that this implies a regression relationship of $\widebar{M}_{j.}$ on $Q_j$, with coefficients $\beta_0$ and $\beta_1$ that are functions of the original model parameters.  Moreover,
\begin{equation}\label{mu_formula}
\mu = \beta_0 + \beta_1 \nu.
\end{equation}
It is easy to derive MLE's for the coefficients in equation \ref{mu_formula}.  The MLE for $\mu$ then follows as
\begin{equation}\label{mu estim}
\widehat{\mu}=\widehat{\beta}_{0}+\widehat{\beta}_{1}\widehat{\nu}.
\;\;\;\   \,
\end{equation}

The MLE's of the regression coefficients and $\nu$ are
\begin{equation*}\label{beta1i}
\widehat{\beta}_{1}=\frac{\sum_{j=1}^{n}\widebar{M}_{j.}Q_{j}-n\widebar{M}^{(n)}\widebar{Q}^{(n)}}{\sum_{j=1}^{n}\left(Q_{j}-\widebar{Q}^{(n)}\right)^{2}} \;\;\; \, ,
\end{equation*}
\begin{equation*}\label{beta0i}
\widehat{\beta}_{0}=\widebar{M}_{j.}-\widehat{\beta}_{1}\widebar{Q}^{(n)} \;\;\;  \, ,
\end{equation*}
and
\begin{equation*}\label{nu}
\widehat{\nu}=\widebar{Q}^{\left(N\right)} \;\;\; \, ,
\end{equation*}
$\widebar{M}_{j.}=K^{-1}\sum_{k=1}^{K}\widebar{M}_{jk}, \widebar{M}^{(n)}=n^{-1}\sum_{j=1}^{n}\widebar{M}_{j.}$ and
$\widebar{Q}^{(n)}=n^{-1}\sum_{j=1}^{n}Q_{j}.$

\end{document}